\documentclass{PoS}

\newcommand{\be}{\begin{equation}}
\newcommand{\ee}{\end{equation}}
\newcommand{\bea}{\begin{eqnarray}}
\newcommand{\eea}{\end{eqnarray}}
\newcommand{\no}{\noindent}
\newcommand{\nn}{\nonumber}

\newcommand{\Tr}{{\rm Tr\,}}
\newcommand{\Det}{{\rm Det\,}}
\newcommand{\e}{{\rm e\,}}

\newcommand{\parm}{\par\medskip}

\newcommand{\rar}{\rightarrow}

\PoS{PoS(LAT2005)170}

\title{The high density region of QCD in a large mass and chemical 
potential model }

\ShortTitle{The high density region of QCD in a large mass and 
chemical potential model }

\author{Roberto De Pietri and Alessandra Feo\thanks{UPRF-2005-03 }\\
        Dipartimento di Fisica, Universit\`a di Parma and INFN Gruppo Collegato di Parma,\\
Parco Area delle Scienze, 7/A, 43100 Parma, Italy\\
        E-mail: \email{depietri@fis.unipr.it} , \email{feo@fis.unipr.it}} 

\author{Erhard Seiler
        \\
        MPI f\"ur Physik - Werner Heisenberg Institut f\"ur Theoretische Physik,
	 M\"unchen, Germany\\
        E-mail: \email{ehs@mppmu.mpg.de}}

\author{\speaker{Ion-Olimpiu Stamatescu}\thanks{I.-O.S. is indebted to 
the MPI M\"unchen for 
hospitality and support while part of
this work has been performed and to DFG for support in attending the conference.}\\
        FEST - Protestant Institute for Interdisciplinary Research, Heidelberg
	and \\ Institut f\"ur Theoretische Physik  der Universit\"at, Heidelberg,
	Germany\\
        E-mail: \email{stamates@thphys.uni-heidelberg.de}}

\abstract{We study the high density region of QCD within an effective model 
obtained in the frame of the hopping parameter expansion. 
The model still acknowledges 
the sign problem peculiar to non-zero chemical potential, but it permits 
the development of refined algorithms which ensure a good overlap of the 
Monte Carlo ensemble with the true one. We review the main features of 
the model, including the most explicit form of the resumed expansion,
 and present calculations concerning
the dependence of various observables on the chemical potential and
on the temperature, in particular of the charge density and the diquark 
susceptibility, which may be used to characterize the various phases 
expected at high baryonic density. }

\FullConference{XXIIIrd International Symposium on Lattice Field Theory\\
		 25-30 July 2005\\
		 Trinity College, Dublin, Ireland}

\begin{document}

\section{Introduction}

The aim of this work is to understand the phase structure of high density,
strongly interacting matter. In the spirit of the $\mu=0$  quenched
approximation one has considered
a ''non-zero density quenched approximation" for $\mu > 0$
based on the double limit $ M \rar \infty,\, \mu \rar \infty,\, \zeta \equiv
{\rm exp}\,(\mu - \ln M) :$ fixed \cite{bend,fktre}. This implements a
static, charged background, which influences the
gluonic dynamics \cite{fktre,bky}. The present model \cite{hdm01} represents 
a systematic extension of the above considerations: 
the gluonic vacuum is enriched by the effects
of dynamical quarks of large (but not infinite) mass, bringing a large net
baryonic charge.  
In \cite{hs} and in 
the present paper we explore the phase structure of the model, 
as a first step in understanding
the properties of such a background. 
 
 This model can be derived as $1/M$ expansion of QCD at large $\mu$
 around the unphysical massive quarks point. However,
 it is more realistic to understand it as an approximation,  
whose justification relies on the predominant role of the gluonic
 dynamics. We want to understand how this 
 dynamics is influenced by the presence of charged matter.
 This would allow, among other things, to  study the effect
 of dense, heavier background baryonic charges on light quarks and hadrons. 
 
 The main ingredient of the model are Polyakov-type of loops, 
 capturing the effect of heavy, limitedly moving  quarks. 
 The model still has a sign problem, but due to the factorization
 of the fermionic determinant, it allows for local, refined 
 algorithms and large statistics.   
 
\section{The model}

The QCD grand canonical partition function with Wilson fermions at $\mu>0$ is  \cite{hkks}:
\bea
&&{\cal Z}(\beta,\kappa,\mu) = \int[DU]\, 
\e^{-S_G(\beta,\{U\})}{\cal Z}_F({ {\kappa}}, \mu, \{U\}), \ \ 
{\cal Z}_F({ {\kappa}}, \mu, \{U\}) =  
\Det W ({ {\kappa}}, \mu, \{U\}), \nn \\
&&W_{ff'} = \delta_{ff'} [ 1 -  \kappa_f\, \sum_{i=1}^3 \left( 
\Gamma_{+i}\,U_i\,T_i +
\Gamma_{-i}\,T^*_i\,U^*_i\right) 
-\kappa_f\,  \left( \e^{\mu_f}\,\Gamma_{+4}\,U_4\,T_4 +
\e^{-\mu_f}\,\Gamma_{-4}\,T^*_4\,U^*_4 \right) ], 
\label{e.act} \nn\\
&&\Gamma_{\pm \mu} = 1 \pm \gamma_{\mu}, \ \ \kappa = 
\frac{1}{2(M+3+\cosh \mu)} = \frac{1}{2(M_0+4)},\,  
\eea
\no with $M$ the ``bare mass", $M_0$: bare mass at $\mu=0$,
$f$: flavour, $U$:   links and $T$: lattice translations.

The hopping parameter expansion is an expansion in closed loops of $\Det W = 
\prod_f \Det W_f$,
\be
\Det W_f = 
{\rm exp} \left[
-\sum_{l=1}^\infty  \sum_{\left\{{\cal
C}_l\right\}} \sum_{s=1}^\infty ~{{{ (\kappa_f^l g^f_{{\cal C}_l})}^s}\over
s}\,\Tr_{\rm D,C}\left({\cal Y}_{{\cal C}_l}{\cal L}_{{\cal C}_l}\right)^s 
\right] 
= 
 \prod_{l=1}^{\infty} \prod_{\left\{{\cal C}_l\right\}} 
\prod_{\alpha=1}^4  
  \Det_{\rm C} \left(1~-~(\kappa_f)^l g^f_{{\cal C}_l}y_{{\cal C}_l}^{\alpha}
{\cal L}_{{\cal C}_l}\right)   \nonumber 
\ee
\be
g^f_{{\cal C}_l}= \left(\epsilon \, \e^{\pm N_{\tau}\mu_f}\right)^r\ {\rm 
if}\ 
{\cal C}_l : ``Polyakov\,\, r\!\! - \!\! path" \,,\ \ 
 =1\  {\rm else}\,,\ {\cal Y}_{{\cal C}_l}= \prod_{l \in {\cal C}_l} 
\Gamma_l\,,\ {\cal L}_{{\cal C}_l} = \prod_{l \in {\cal C}_l} U_l 
 \label{e.hopg}
\ee
\no where ${\cal C}_l$ are distinguishable, 
non-exactly-self-repeating 
closed paths of length $l$, $s$ is the number of 
 coverings of  ${\cal C}_l$,
and a $``Polyakov\ r\!\! - \!\! path"$  winds $r$ times
over the lattice 
in the $\pm 4$ direction with
periodic(antiperiodic) b.c. $\epsilon = +1(-1)$ (p.b.c. in spatial directions).
The factors $y_{{\cal C}_l}^{\alpha}$ are the eigenvalues of ${\cal Y}_{{\cal C}_l}$
 and the expansion  can be expressed as a product of colour determinants \cite{sdet}. 
Notice that $\Det W$ is a polynomial in $\kappa$ of the order $4\,n_c\,n_f\,N_V$ 
(colours, flavours, lattice volume), which means that the 
expansions on the right hand side can be truncated at this order.
 
\begin{figure}[t]
\vspace{4.5cm} 
\includegraphics{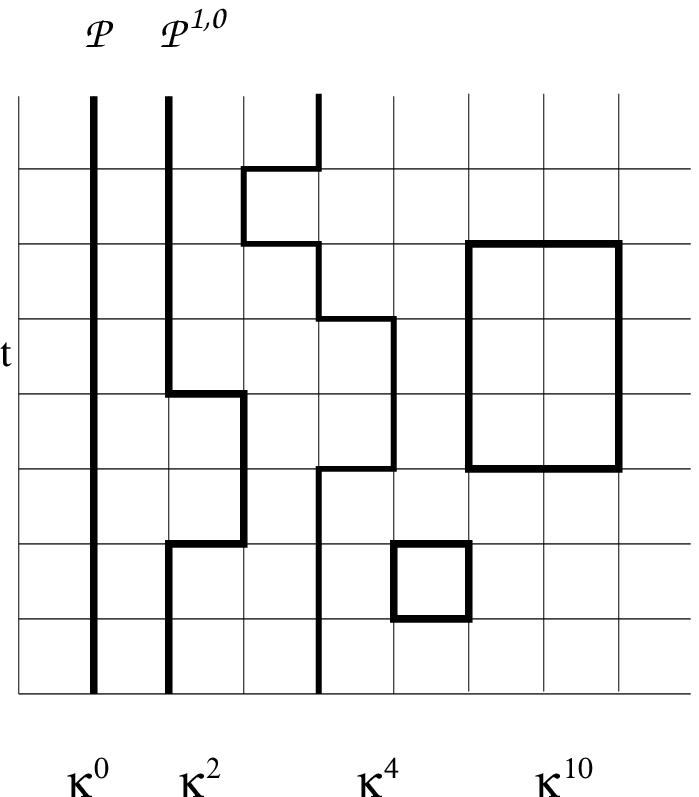} 
\includegraphics{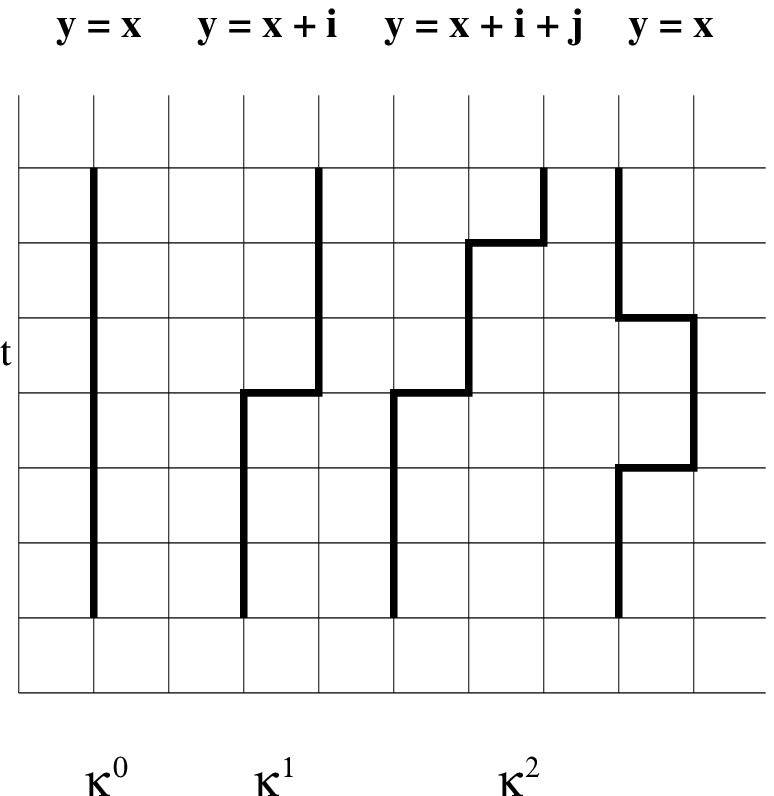} 
\includegraphics{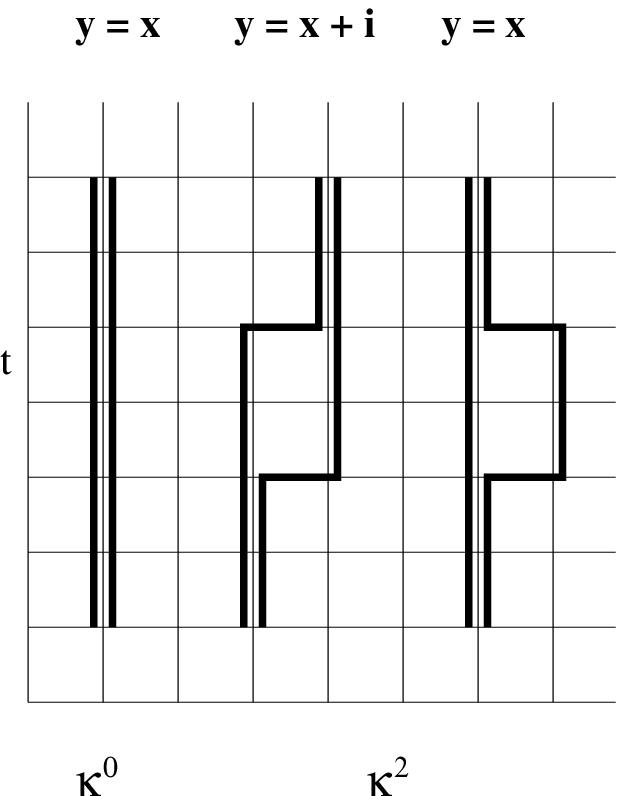} 
\caption{Loops in the determinant (left); quark (middle) and di-quark propagators (right).
} 
\label{f.qq} 
\end{figure}

The ``quenched limit at $\mu>0$" is \cite{bend}:
\bea
&&\kappa \rightarrow 0\,,\ \mu \rightarrow \infty\,,\ \ 
\kappa\, \e^{\mu} \equiv \zeta:\ {\rm fixed}\,, \  C = (2\, \zeta)^{N_\tau}
\label{e.qul}\\
&&{\cal Z}_F^{[0]}(C, 
\left\{U\right\}) = \exp \left[- 
  \sum_{\left\{{\vec x}\right\}}
\sum_{s=1}^\infty\!\! ~{{{ (\epsilon C)}^s}\over s}~\Tr_{\rm C} 
  ({\cal P}_{\vec x})^s \right] 
=\, \prod_{\left\{{\vec x}\right\}}~ 
  \Det_{\rm C} \left(1~-~\epsilon\,C {\cal P}_{\vec x}\right)^2,
\,\,\,\,\,\, \label{e.hdl} 
\eea 
\no and the next order corrections  lead to \cite{hdm01}:
 \bea
{\cal Z}_F^{[2]}({ {\kappa}}, \mu, \left\{U\right\}) 
 &=&   {\rm exp}\left\{-2\,  \sum_{\left\{{\vec x}\right\}}\,
\sum_{s=1}^\infty \,{{{ (\epsilon\, C
)}^s}\over s} \, \Tr_{\rm C} 
  \left[({\cal P}_{\vec x})^s  + \kappa^2\sum_{r,q,i,t,t'}
(\epsilon\, C)^{s(r-1)}({\cal P}_{{\vec x},i,t,t'}^{r,q})^s \right]\right\}
\nn\\
&=&  {\cal Z}_F^{[0]}( C,  \left\{U\right\}) \,  
 \!\!\!\!\!\!  \prod_{{\vec x}, r,q,i,t,t'} 
  \Det_{\rm C} \left(1-(\epsilon\,C)^{r}\,\kappa^2\,
  {\cal P}_{{\vec x},{i},t,t'}^{r,q}\right)^2 . 
\label{e.corr2}
\eea
See Fig. \ref{f.qq}. For easy bookkeeping we use the incomplete temporal gauge:
\bea
\!\!\!\!\!\! U_{n,4}=1 &,& {\rm except}\ U_{({\vec x}, n_4=N_\tau),4} \equiv V_{\vec x}\,,
\label{e.ptg}\\
\!\!\!\!\!\! {\cal P}_{\vec x}= V_{\vec x}&,& {\cal P}_{{\vec x},i,t,t'}^{r,q}= (V_{\vec x})^{r-q} 
U_{({\vec x},t),i} (V_{{\vec x}+{\hat {i}}})^q 
U_{({\vec x},t'),i}^*\,,\  r>q\ge 0,\, \pm i =1, 2, 3, \,
1\le t \le t' \le N_{\tau}\, \ \label{e.pltg}
\eea
\no ($t < t'$ for $q=0$). Eqs. 
(\ref{e.qul}-\ref{e.pltg}) define the present model. Notice that for $U \in$ SU(3) we
have:
\be
\Det_{\rm C} \left(1~-~\epsilon\,C U\right)=
1 - \,\epsilon\,C \Tr U+\,C^2\,\Tr U^{\dagger} - \epsilon\,C^3
\ee
\no We measure spatial and temporal plaquettes, 
Polyakov loops and baryon charge densities
\bea
\frac{n_B}{T^3}=\frac{N_{\tau}^3}{3 N_{\sigma}^3} {\hat n}, \ \
{\hat n}  =   \hat n_0 +\hat n_1,\ \ 
\hat n_0 = \langle \frac{\partial}{\partial \mu}{\ln \cal Z}_F^{[0]} \rangle 
, \ \  
\hat n_1 =  \langle \frac{\partial}{\partial \mu}
\left(\ln \frac{{\cal Z}_F^{[2]}}{{\cal Z}_F^{[0]}}\right) \rangle 
\eea
\no  as well as 
diquark propagators (in maximal temporal gauge) and the
corresponding susceptibility, which could
give information on the large $\mu$ physics \cite{arw} (see \cite{hs}
for explicite formulae, see Fig. \ref{f.qq}).

\section{Algorithm and Simulation}

 We use the Wilson action and Wilson fermions within a
reweighting procedure.
The updating is done with a local Boltzmann factor (which 
only leads to a redefinition of the 
``rest plaquette"): 
\bea
\prod_{Plaq}\e^{\frac{\beta}{3}\,\Tr\,Plaq}\,
\prod_{\vec x}
{\rm exp}\left\{ 2\,C\,
{\cal R}e\Tr_{\rm C} 
  \left[{\cal P}_{\vec x}  + \kappa^2\sum_{i,t,t'}
{\cal P}_{{\vec x},i,t,t'}^{1,0} \right]\right\} \,. \label{e.bolr}
\eea
We here employ the Cabbibo-Marinari heat-bath procedure
mixed with over-relaxation. This updating already takes into account 
part of the $\mu>0$ effects and the generated ensemble
can thus have a better overlap with the true one than an
 updating at $\mu=0$. Notice that such a factor may also improve 
  convergence of full QCD simulations at $\mu>0$.  The
 weight (global, vectorizable) is:
\bea
\prod_{\vec x}
{\rm exp}\left\{ -\,2\,C\,
{\cal R}e\Tr_{\rm C} 
  \left[{\cal P}_{\vec x}  + \kappa^2\sum_{i,t,t'}
{\cal P}_{{\vec x},i,t,t'}^{1,0} \right]\right\} 
{\cal Z}_F^{[2]}(\left\{U\right\}). \label{e.wght}
\eea
\no One also can use an improved partition between the updating
factor and the weight, to be taken care of by a supplementary
Metropolis check. Anisotropy can be straightforwardly  introduced.

The simulations are mainly done on lattices  $6^4$ and $8^4$
(in the incomplete temporal gauge (\ref{e.ptg})). 
We shall present results for 
$n_f=1$ and  $n_f=3$ degenerate flavours 
(any mixture of flavours can be
implemented).
The $\kappa$ dependence 
has been analyzed in \cite{hdm01}. Here we set $\kappa=0.12$,
which drives the $1/M^2$  effects  in the baryonic density to
 about $50$\%.
Our problem setting is primordially to explore the
 phase structure of the model at large $\mu$, small
$T$ and we accordingly vary $\beta$ and $\mu$.

\section{Results}

\begin{figure}[hb]
\vspace{5.2cm} 
\includegraphics{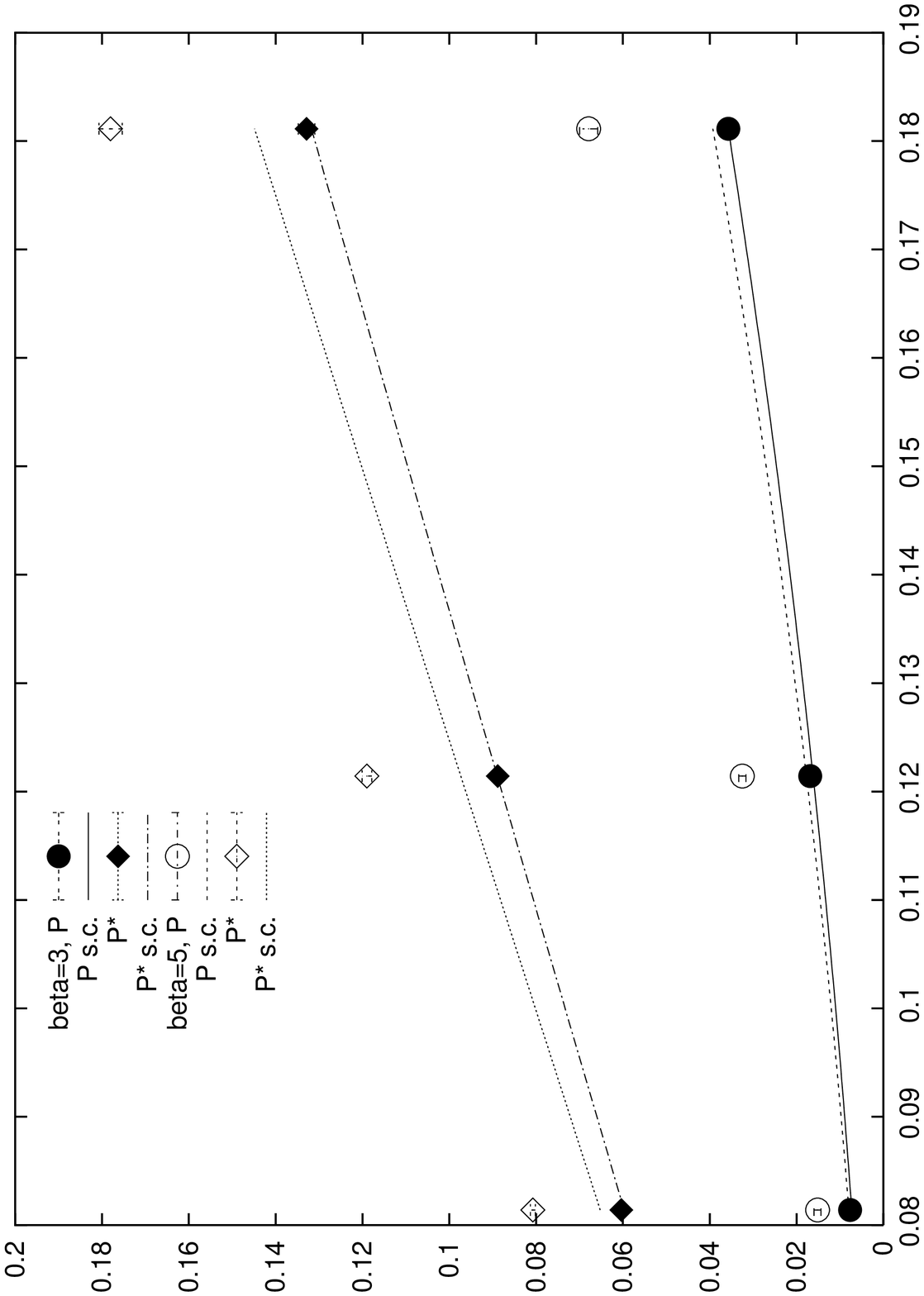} 
\includegraphics{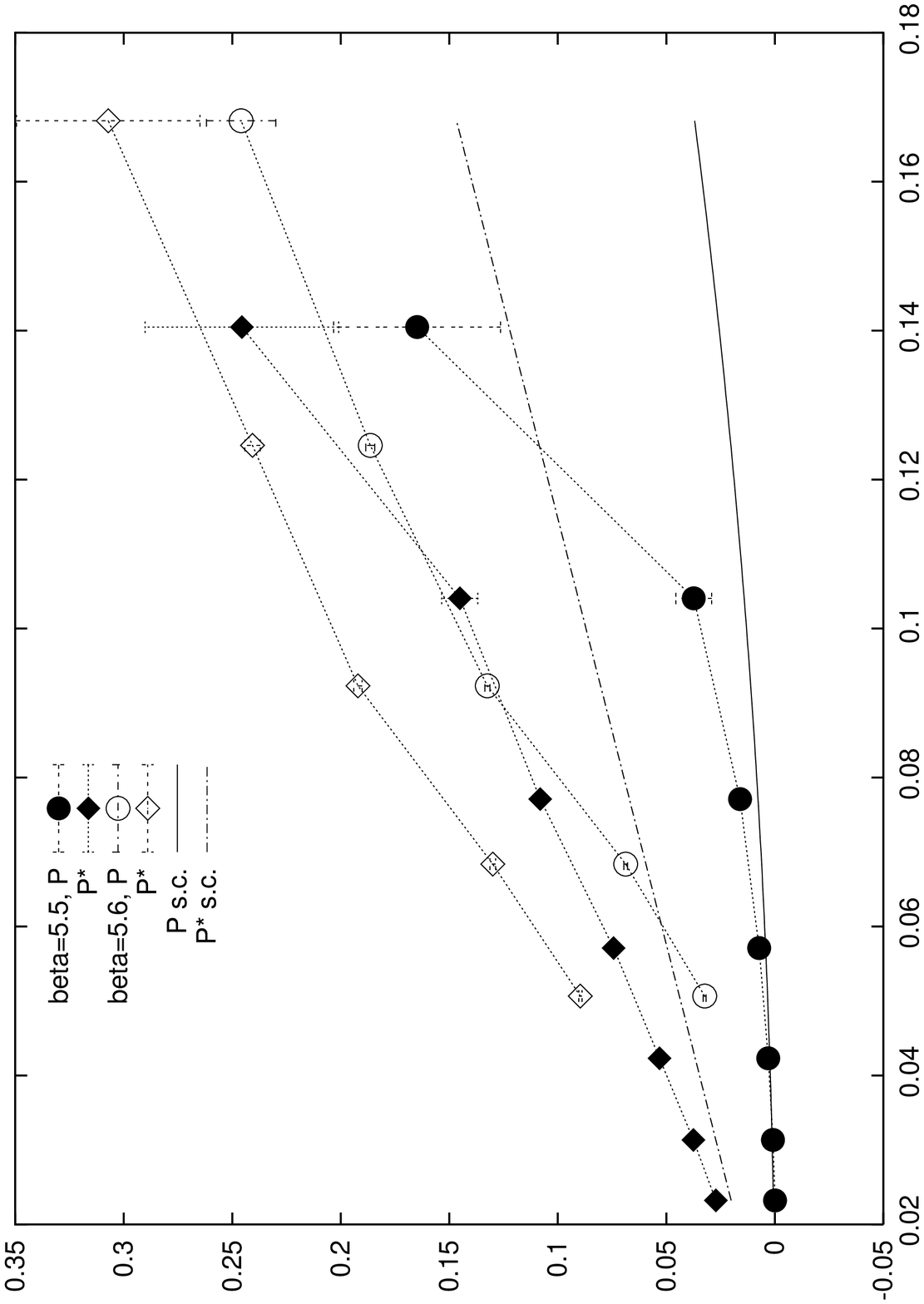} 
\caption{Polyakov loops $P$ (triangles) and $P^\dag$ (diamonds) {\it vs} $C$, 
$4^4$ (left, $\beta=3$, full  and $\beta=5$, open symbols) 
and $6^4$ (right, $\beta=5.5$, full symbols and $\beta=5.6$, open symbols),
 strong coupling (lines).} 
\label{f.stro} 
\end{figure}

Some analytic results can be obtained in strong coupling. 
To lowest order in $\beta$ we have: 
\bea
P &=&\frac{1}{3} {\rm Tr}
\langle \frac{1}{N_{\sigma}^3}\sum_{\vec x} \Tr {\cal P}_{\vec x} \rangle
= C^2\,\frac{1+\frac{2}{3}C^3}{1+4C^3+C^6}\,\left[1 +
\frac{2N_\tau}{3}\,\beta\,\kappa^2\,\frac{1+ 3C^2+2C^3 + 2C^5 +C^6}
{1+\frac{2}{3}C^3}\right]\, \\
P^{\dag} &=&\frac{1}{3} {\rm Tr} 
\langle \frac{1}{N_{\sigma}^3}\sum_{\vec x} \Tr {\cal P}_{\vec x}^{\dag} \rangle 
=\frac{2}{3}C\,\frac{1+C^3}{1+4C^3+C^6}
\,\left[1 +
\frac{2N_\tau}{3}\,\beta\,\kappa^2\,\frac{1+ 3C^2+2C^3 + 2C^5 +C^6}
{1+C^3}\right]\,
\eea
\no  for $n_f=1$.  Fig. \ref{f.stro} shows increasing
 deviation of the full results from strong coupling with increasing $\beta$
and indication of qualitative change of behaviour at large $\mu$ and above
$\beta=5.5$.

\begin{figure}[ht]
\vspace{5.2cm} 
\includegraphics{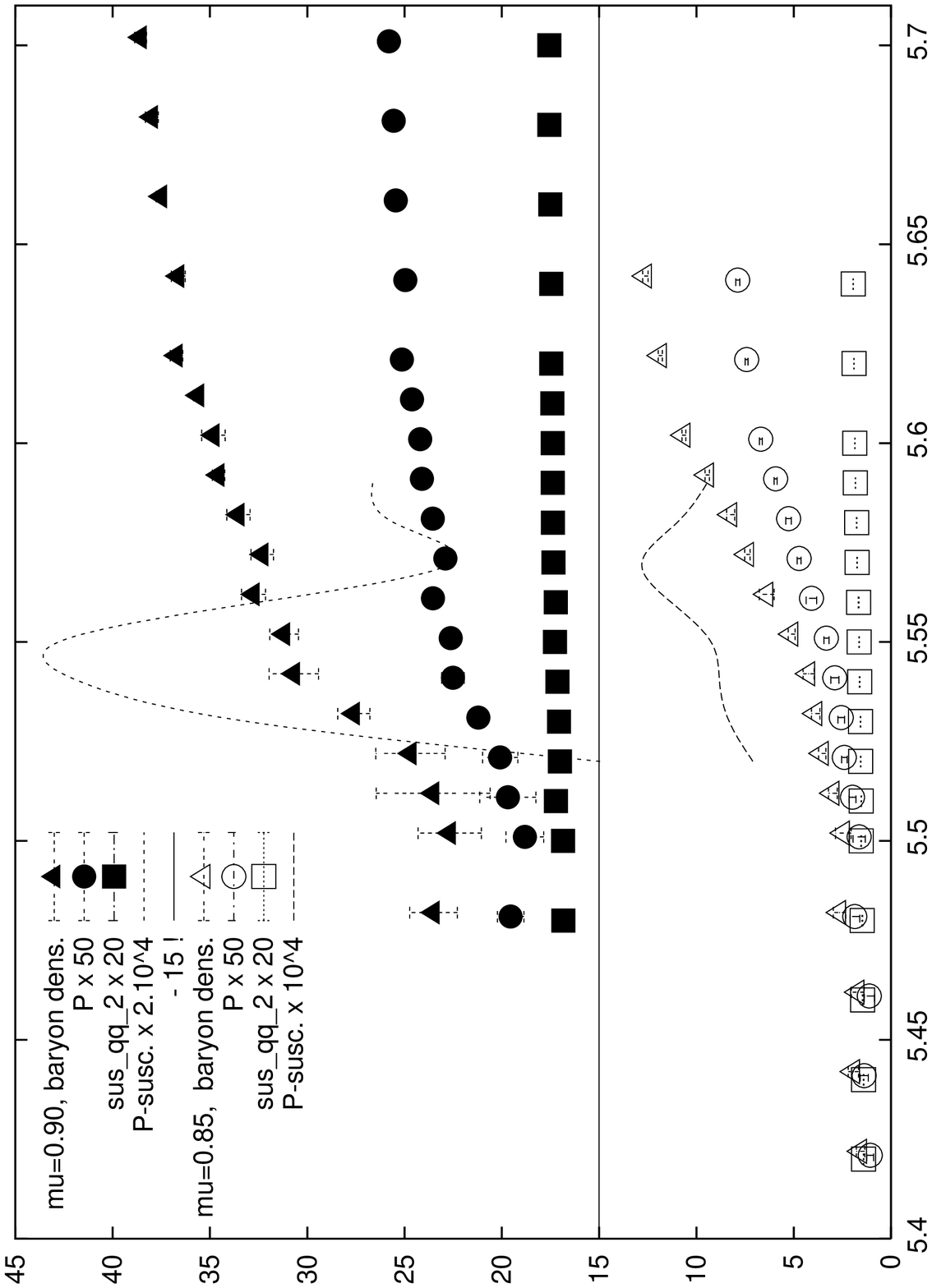} 
\includegraphics{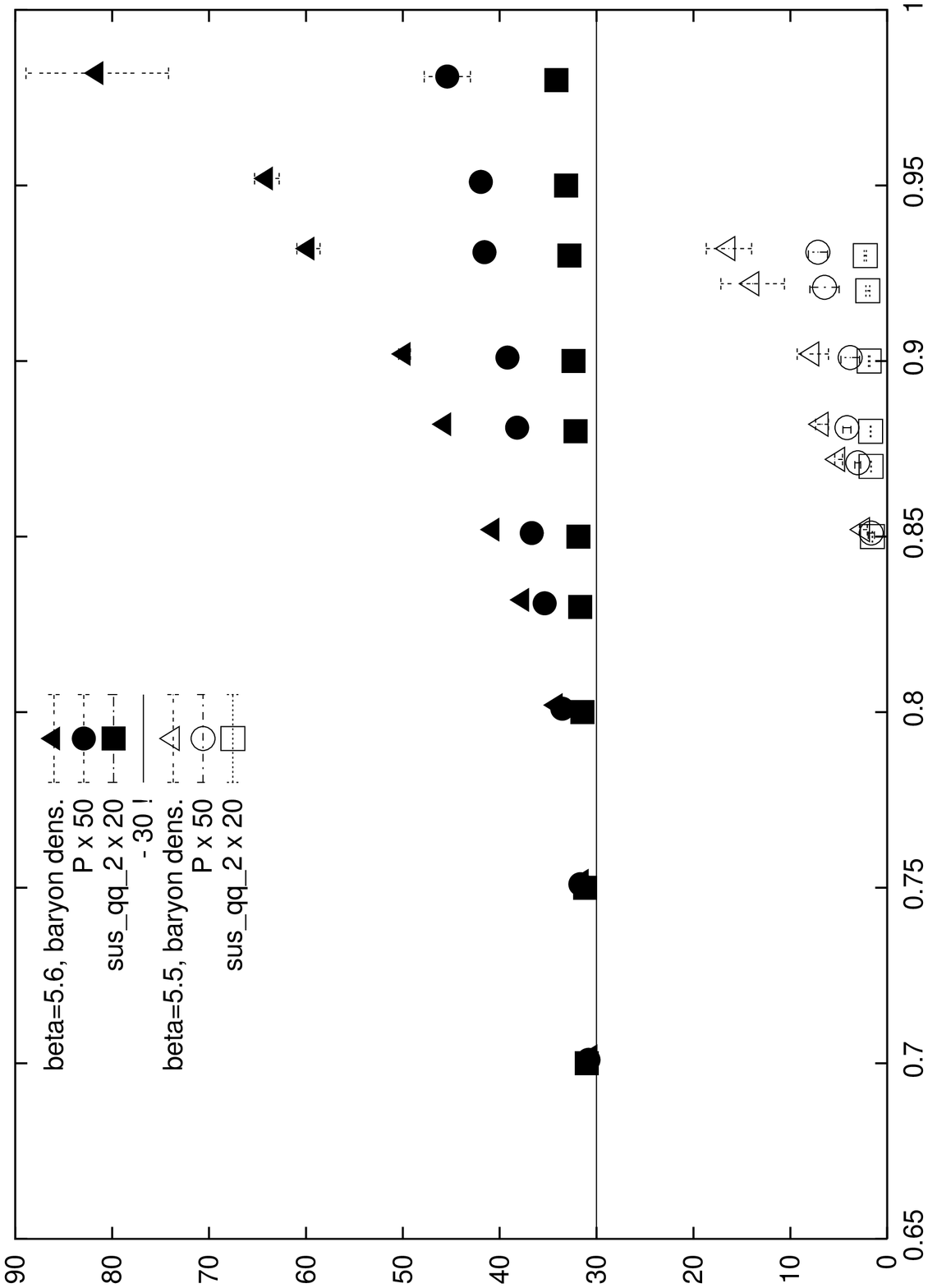} 
\caption{Baryonic density (triangles), Polyakov loops (circles) and 
diquark susceptibility (squares) {\it vs}
$\beta$ (left, $\mu=0.85$ open, and $\mu=0.90$ full symbols) and {\it vs} $\mu$
 (right, $\beta=5.5$ full and $\beta=5.6$ open symbols).} 
\label{f.dbmu} 
\end{figure}

On Fig. \ref{f.dbmu}, left plot, we show for $n_f=3$ the behaviour with 
$\beta$ at fixed $\mu$ values, which in the  physical $T,\mu$ plane would mean
moving on lines  $T \simeq \frac{1}{\mu N_{\tau}}\mu_{\rm phys}$. 
 We see at both values of $\mu$ 
 qualitative changes of behaviour suggesting transitions
from low to high temperature phases at values of $\beta$ as indicated by the 
Polyakov loop susceptibility
(curves on the plot). 
 This appears corroborated by the behaviour shown on
the right hand plot, where we vary $\mu$ at two fixed $\beta$ values and see a 
delayed onset
of the ``transition" for the smaller $\beta$. 
The diquark susceptibility (we only show the second order term) is
rather flat in $\beta$ and shows a signal only at  $\mu$ approaching $1$.

\begin{figure}[ht]
\vspace{5.2cm} 
\includegraphics{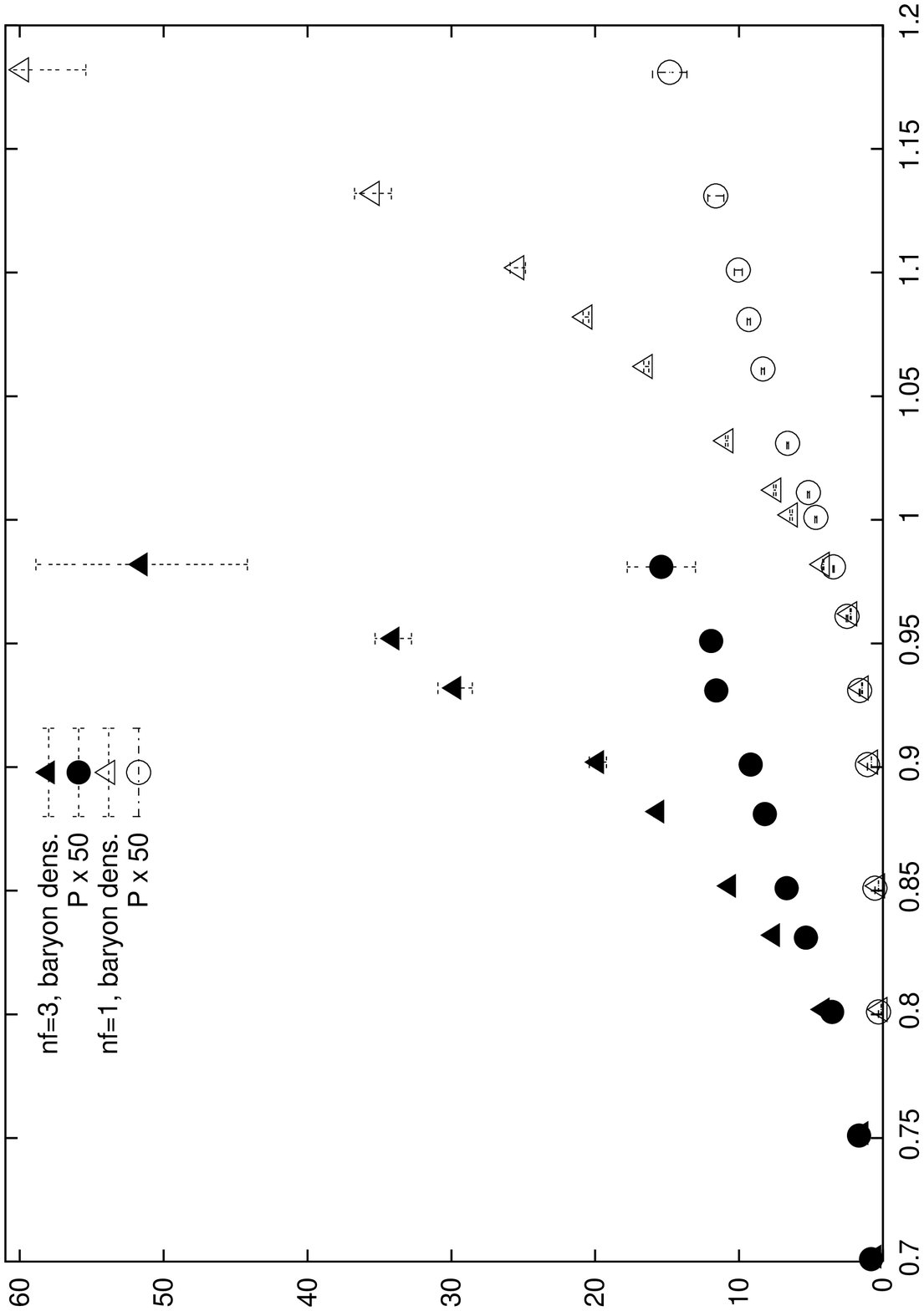} 
\includegraphics{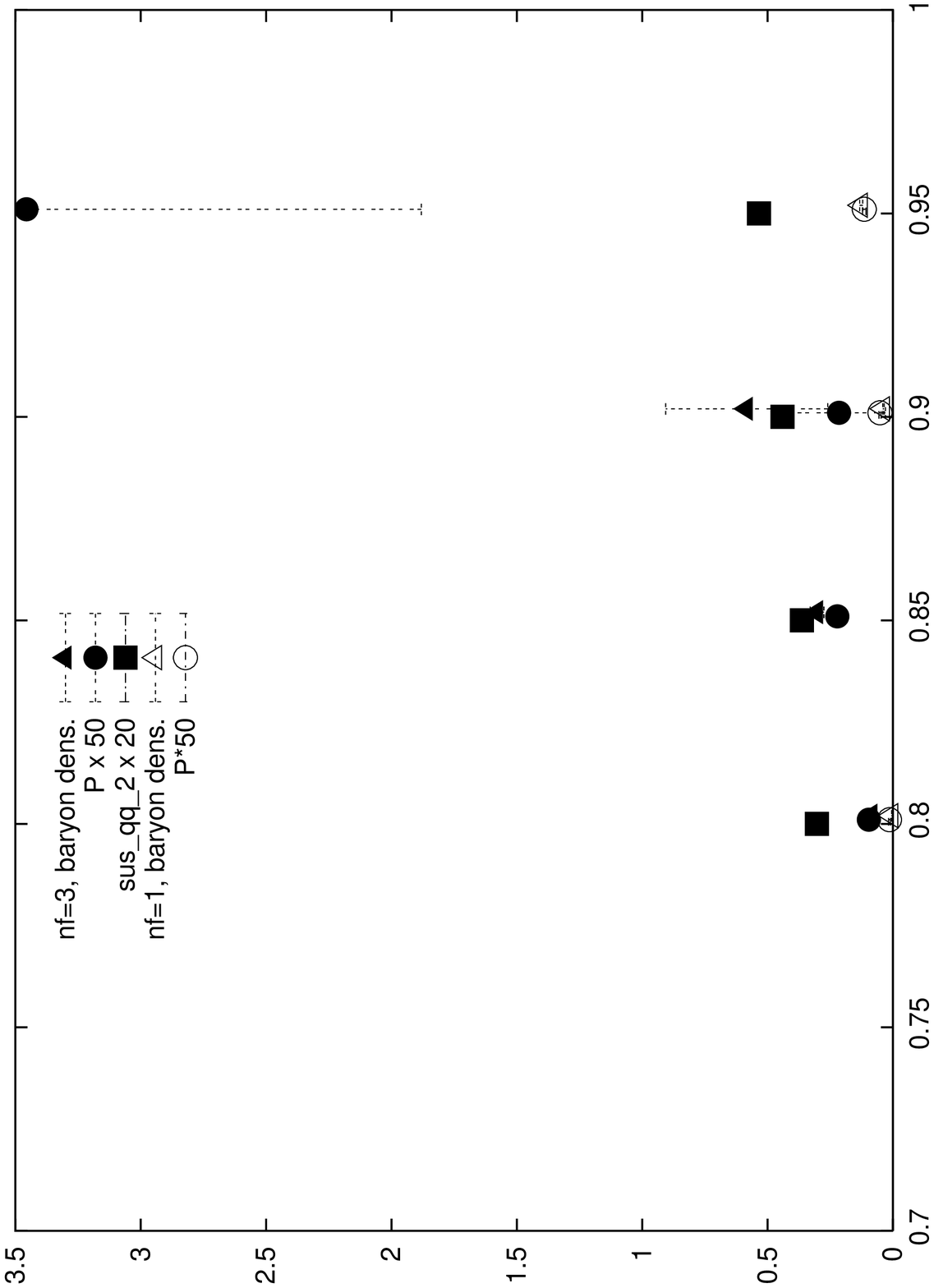} 
\caption{Baryonic density (triangles) and Polyakov loops (circles) {\it vs}
$\mu$ at $\beta=5.6$,  $6^4$ (left) and $8^4$ (right; here also the diquark
susceptibility, squares): $n_f=1$ open and 
$n_f=3$ full symbols.}
\label{f.demu} 
\vspace{-0.5cm} 
\end{figure}

The dependence on $n_f$ is illustrated on Fig. \ref{f.demu}, both for the 
$6^4$ and the $8^4$ lattices. As one can see, 
 the highest values of $\mu$ attainable in this method are about $5 \sim 6\, T$,
 depending on $n_f$.

\section{Discussion}

\begin{figure}[ht]
\vspace{5.2cm} 
\includegraphics{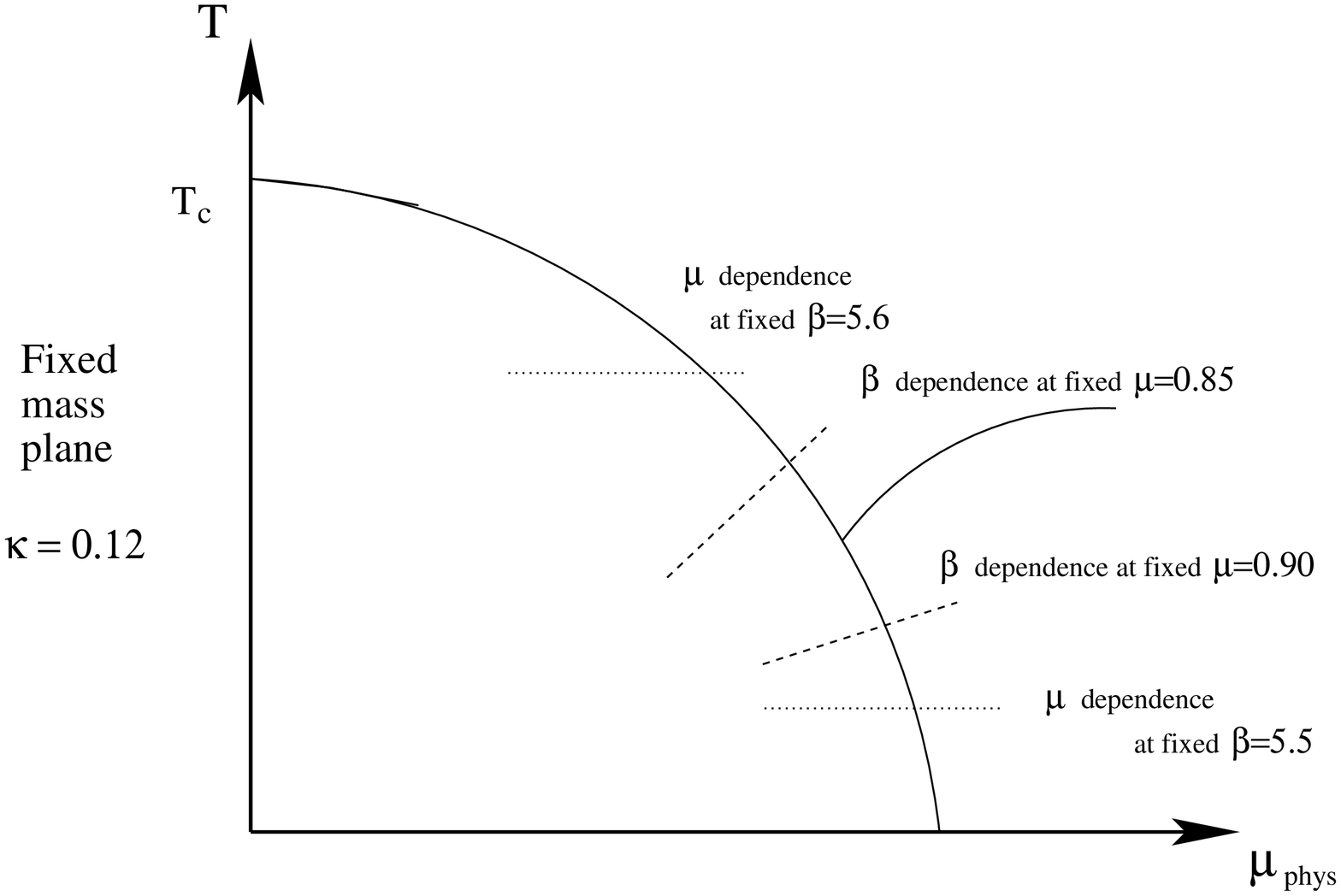} 
\includegraphics{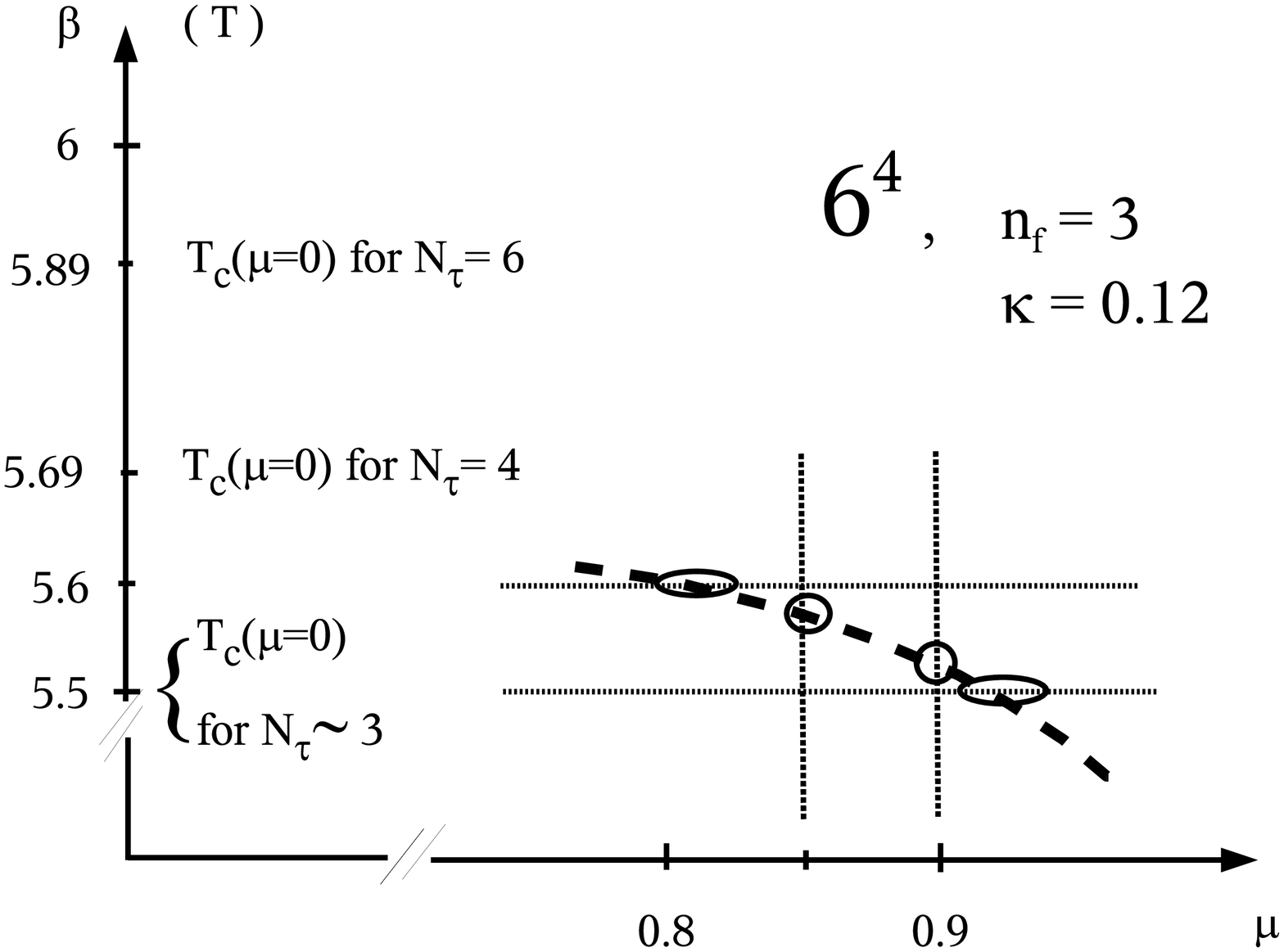} 
\caption{{Prospected phase structure (only indicative!) and observed changes 
of behaviour (transitions?).}
}
\label{f.phd} 
\vspace{-0.3cm} 
\end{figure}

Our calculations show strong effects at large $\mu$, which even at
moderate $\beta$ depart considerably from strong coupling estimates
and also indicate possible phase transitions.
The results concerning the latter for the $n_f=3$ case
are summarized in Fig. \ref{f.phd}. On the left plot we indicate
the lines we have followed in the simulation in the $T,\mu_{\rm phys}$ plane.
On the right hand plot we show the $\beta,\mu$ plane together with the points at which 
we see indication of transitions. Since we cannot yet fix a scale (but see below) we 
preferred to show these points in the bare parameter plane. While the indication
 for a
transitions are clear, it is uncertain what
happens at larger $\mu$ and small $T$ and how the transition lines run.
This region needs therefore further study.

The algorithm works reasonably well over a wide range of parameters
and for lattices up to $8^4$.
We reach large densities and $\frac{\mu}{T}$ for temperatures $\sim 
\frac{1}{3}\,T_c(\mu=0)$ or less.
It appears difficult, however, to go to larger lattices and larger $\mu$
with this algorithm and
one should consider improving it.

The model permits to vary $\mu$, $\kappa$ and $T$ as 
independent parameters. It is therefore interesting to extend the study
to cover this full variability. For higher orders in $\kappa$ the bookkeeping
soon becomes unmanageable, one could however consider using statistical
ensembles of large loops \cite{mn}.

Concerning the significance of this analysis
we can take two points of view:

Firstly, we can consider this model for itself, as describing  
``quasi-static charges" interacting via gauge forces.  
One may then ask  whether this dynamics may lead to a non-trivial 
phase structure.

Secondly, we can consider this model as an evolved ``quenched approximation" in 
 the presence of charged matter. Then this study would give us information
about the structure of the so modified gluonic vacuum of the SU(3) theory. It would then be
 natural to think of it as providing  
 a heavy, dense, charged
background for light quarks propagation and calculate light
hadron spectra and other hadronic properties under such conditions. 
This could also help fixing a  scale.\parm

Acknowledgments: The calculations have been done on the VPP5000 computer of the University of
Karlsruhe and on the PC Cluster of the Institute of Physics of the Parma
University.

\end{document}